\documentclass[a4paper,11pt]{amsart}
\begin{document}
\hyphenation{gra-vi-ta-tio-nal re-la-ti-vi-ty Gaus-sian
re-fe-ren-ce re-la-ti-ve gra-vi-ta-tion Schwarz-schild
ac-cor-dingly gra-vi-ta-tio-nal-ly re-la-ti-vi-stic pro-du-cing
de-ri-va-ti-ve ge-ne-ral ex-pli-citly des-cri-bed ma-the-ma-ti-cal
de-si-gnan-do-si coe-ren-za pro-blem gra-vi-ta-ting geo-de-sic
per-ga-mon ina-de-qua-te Fe-brua-ry phy-si-cally}
\title[GW's towards fundamental principles of GR]
{{\bf GW's towards fundamental principles of GR}}

\author[Angelo Loinger]{Angelo Loinger}
\address{A.L. -- Dipartimento di Fisica, Universit\`a di Milano, Via
Celoria, 16 - 20133 Milano (Italy)}
\email{angelo.loinger@mi.infn.it}

\vskip0.50cm

\begin{abstract}
The physical non-existence of gravitational waves (GW's) as a
consequence of the non-existence in general relativity (GR) of
physically privileged reference frames, and of the ``plasticity''
of relativistic notion of a coordinate system.
\end{abstract}

\maketitle


\\ PACS 0.40.30 --
Gravitational waves and radiation: theory.

\normalsize

\vskip1.20cm \noindent \textbf{1}. -- In 1944 Weyl proved that the
\emph{linearized} approximation of GR is
\textbf{\emph{inadequate}} to an appropriate treatment of the
so-called GW's \cite{1}.

\par Moreover, in the \emph{exact} (non-approximate) formulation
of GR there exist various demonstrations of the \emph{physical}
non-existence of GW's \cite{2}.

\par In this Note I give another proof that the GW's are only
mathematical undulations quite destitute of a physical reality.
The argument is composed of two propositions, that have been
already sketched by me in different contexts. However, for its
significance, the question deserves a more complete illustration.

\par The following proof is qualitative and very simple. It is
a straightforward consequence of the freedom of choice of the
reference systems and of their ``plasticity'' \cite{3}.

\vskip0.80cm \noindent \textbf{2}. -- In the \emph{exact} GR we
have no class of physically privileged reference systems.  Now,
consider the spacetime manifold $V$ connected with a celestial
body $B$, and an observer $\Omega$ who is at rest together with
$B$; $\Omega$ does not register any gravitational undulation
emitted by $B$. (In Maxwell theory an inertial observer $I$, who
is moving together with a charge $C$, does not register any e.m.
wave emitted by $C$). However, the same conclusion holds also for
a distant observer $\Omega'$, who is in motion with respect to
$\Omega/B$: for him too, $B$ does not send forth any GW. (In
Maxwell theory an inertial system $I'$, who is in motion with
respect to $I/C$, does not register any e.m. wave emitted by $C$).
Indeed, if $\Omega'$ found that $B$ has emitted a GW, we could
affirm that reference $\Omega'$ has a privilege which is not
shared by reference $\Omega$, \emph{i.e.} that $\Omega'$ belongs
to a privileged class of reference frames.

\par Furthermore, as it was remarked by Weyl \cite{3}, ``\ldots \emph{der Begriff
der Relativbewegung zweier K\"{o}rper gegeneinander in der
allgemeinen Relativit\"{a}tstheo\-rie ebensowenig einen Sinn hat
wie der Begriff der absoluten Bewegung eines einzigen}. Solange
man noch den starren Bezugsk\"{o}rper zur Verf\"{u}gung hatte und
an die Objektivit\"{a}t der Gleichzeitigkeit glauben konnte, auf
dem Standpunkte Machs etwa, unter der Herrschaft der
'kinematischen Gruppe' gab es eine relative Bewegung; aber in der
allgemeinen Relativit\"{a}tstheorie hat sich das Koordinatensystem
so 'erweicht', da\ss{} auch davon nicht mehr die Rede sein kann.
Wie die beiden K\"{o}rper sich auch bewegen m\"{o}gen, immer kann
ich durch Einf\"{u}hrung eines gleigneten Koordinatensystems sie
beide zusammen auf Ruhe transformieren.'' \emph{I.e.}: in GR the
notion of reference frame has acquired a characteristic
``plasticity'', so that it is always possible to choose a
coordinate system for which, \emph{e.g.}, \emph{both} $\Omega /B$
and $\Omega'$ are \emph{at rest}. Another application of Weyl's
remark: the orbital motions of the stars of a binary system can be
``obliterated'' with a suitable choice of the coordinates -- with
the obvious conclusion that their revolution motions cannot give
out any GW; \emph{a conspicuous instance is represented by the
famous binary} B PSR1913+16. (Clearly, the reductions to rest are
possible owing to the ``transport, or inertial, forces''
associated with the convenient coordinate frames).

\vskip0.80cm \noindent \textbf{3}. -- It is instructive to recall
that similar, but less general, results exploiting the
``plasticity'' of the coordinate systems had been obtained in 1924
by Hilbert \cite{4} and in the Sixties of past century by Landau
and Lifchitz \cite{5}.

\par Hilbert emphasizes that in the problem of the Einsteinian
field of a gravitating  body (extended or point-like) of a given
mass $M$, we can choose a coordinate system for which a moving
test-particle and the gravitating body are \emph{both} reduced to
rest. He gives a simple example of this fact. To be determinate
and physically significant, assume that our gravitating body has
the minimal radius $r_{min}=(9/8)(2m)$, where $m\equiv GM/c^{2}$.
Let us consider a circular motion on an orbit
$r=\textrm{constant}$. Hilbert proves that this motion is
\emph{restricted} by the following conditions (Hilbert puts $c=1$;
his $\alpha$ is equal to $2m$):

\begin{equation} \label{eq:one}
r> \frac{3}{2} \,(2m) >  \frac{9}{8} \,(2m)\quad,
\end{equation}

\begin{equation} \label{eq:two}
v < \frac{1}{\sqrt{3}} \quad,
\end{equation}

where $v=(m/r)^{1/2}$ is the linear velocity.

\par For $r=r_{0}$, say, the angular velocity
$(\textrm{d}\varphi/\textrm{d}t)_{0}$ is given by

\begin{equation} \label{eq:three}
\left(\frac{\textrm{d}\varphi}{\textrm{d}t}\right)_{0} =
\left(\frac{m}{r_{0}^{3}}\right)^{1/2} \quad .
\end{equation}

 \par If we consider a co-rotating reference system, the centre of
 which coincides with the centre of the gravitating body, we have:

\begin{equation} \label{eq:four}
\varphi \rightarrow \varphi +
\left(\frac{m}{r_{0}^{3}}\right)^{1/2} t \quad;
\end{equation}

thus, the spacetime interval

\begin{equation} \label{eq:five}
\frac{r}{r-2m} \, \textrm{d}r^{2} + r^{2} \, \textrm{d}\varphi^{2}
- \frac{r-2m}{r} \, \textrm{d}t^{2}
\end{equation}

is transformed into
\begin{equation} \label{eq:six}
\frac{r}{r-2m} \, \textrm{d}r^{2} + r^{2} \, \textrm{d}\varphi^{2}
+2 \left(\frac{m}{r_{0}^{3}}\right)^{1/2} \,
r^{2}\textrm{d}\varphi \, \textrm{d}t +
\left(\frac{m}{r_{0}^{3}}r^{2} - \frac{r-2m}{r} \right) \,
\textrm{d}t^{2} \quad.
\end{equation}

\par For $r=r_{0}$, we have:

\begin{equation} \label{eq:seven}
\frac{r_{0}}{r_{0}-2m} \, \textrm{d}r^{2} + r_{0}^{2} \,
\textrm{d}\varphi^{2} +2 \left(m \cdot r_{0} \right)^{1/2} \,
\textrm{d}\varphi \, \textrm{d}t + \left(\frac{3m}{r_{0}} -1
\right) \, \textrm{d}t^{2}\quad,
\end{equation}

and since $r_{0}> (3/2)(2m)$, the \emph{pseudo}-Riemannnian
property of the interval is preserved, \emph{i.e.} transformation
(\ref{eq:four}) is an appropriate spatiotemporal transformation,
which reduces actually to rest \emph{both} the test-particle and
the gravitating body.

\par Of course, Hilbert adopts the \emph{standard} form of
solution to Schwarzschild problem, which was discovered by him, by
Droste, and by Weyl, independently. And it is very interesting
that inequality (\ref{eq:one}) is stronger than inequality $r>2m$,
which is the \emph{validity condition} of the standard form -- as
it was prescribed by its discoverers.

\par The inequalities (\ref{eq:one}) and (\ref{eq:two}) are quite
``anti-Newtonian'': they are a consequence of the fact that, for
suitably small values of $r$, the spacetime curvature exerts on
the test-particle a \textbf{\emph{repulsive}} action. And we see
that the pseudo-Riemaniann character of interval (\ref{eq:seven})
is closely connected with this gravitational repulsion.

\vskip0.50cm \noindent \textbf{3bis}. -- Landau and Lifchitz
\cite{5} observed that the maximal number of test-particles moving
in a given gravitational field, which can be reduced to rest by
means of a convenient choice of coordinates, is \emph{four}.
Precisely, if we assume that the particles are situated at the
four vertices of a tetraedron (which has six edges), it is
possible to find a coordinate system such that they are at rest.
Obviously, the choice of a tetraedron is not casual; in
particular, just six are the essential components of the metric
tensor.

\vskip0.50cm \noindent \textbf{3ter}. -- The world-line of any
test-particle $P$ in a given gravitational field is a
\emph{geodesic} line, and therefore no emission of GW's is
possible. The reduction to rest of $P$ yields a reinforcement of
this conclusion.

\vskip0.50cm \noindent \textbf{4}. -- GR and Maxwell e.m. theory
are basically different: in Maxwell theory the inertial frames are
physically privileged, and the accelerated charges emit e.m.
waves; in GR, on the contrary, \emph{no} reference system is
physically privileged, and the accelerated masses do not emit
GW's.

\par The formal (and partial) resemblance of the \emph{linear}
version of GR with the e.m. theory has deceived many physicists.
Unfortunately, the important paper by Weyl of 1944 \cite{1} is
ignored in the current literature.

\vskip0.50cm \noindent \textbf{5}. -- Another consequence of the
``plasticity'' of the coordinate systems is the following (cf.
\cite{6}). Let us consider a swarm of corpuscles, gravitationally
and non-gravitationally interacting, and fix our attention upon
the world-line $L$ of any of them. By virtue of a beautiful and
well-known theorem by Fermi, we can choose a coordinate system for
which the metric tensor is \emph{constant} on the whole line $L$.
This means that the gravitational field \emph{on $L$} has been
blotted out. Consequently, no GW has been emitted by our
corpuscle. Now, an identical argument can be applied to the other
corpuscles, singularly taken into consideration. (If the
corpuscles interact only gravitationally, their world-lines are
\emph{geodesic} lines).

\vskip0.50cm \noindent \textbf{6}. -- A last and impressive
consequence of the ``plasticity'' of the coordinate systems in GR:
it is easy to see that the \emph{undulatory character} of any wavy
solution of Einstein field equations can be crossed out by a
suitable change of the coordinates. The result is a metric tensor
which is generally non-stationary.

\begin{center}
\noindent \small \emph{\textbf{APPENDIX}}
\end{center}

\normalsize \noindent \vskip0.20cm
\par In February 2007 and in March 2007 the LIGO Scientific
Collaboration published two papers, resp. entitled: ``Upper limits
on gravitational wave emission from 78 radio pulsars'' \cite{7},
and ``Search for gravitational wave radiation associated with the
pulsating tail of the SGR 1806-20 hyperflare of 27 December 2004
using LIGO'' \cite{8}.

\par The apparatuses did not register any GW.
\par \emph{A comment}. The LIGO researchers have an unjustified
belief in the validity of the obsolete theory of GW's,  which is
derived from the linearized approximation of GR, and in the
physical adequacy of the various undulatory solutions of Einstein
field equations (see, \emph{e.g.} \cite{9}), which have a mere
formal character, \emph{i.e.} a character devoid of physical
significance \cite{2}. These unfounded convictions induce them to
persevere in the chase of a non-existing phenomenon: the GW's.

\par (In 1977 D. Kennefick wrote a review article entitled ``Controversies
in the History of the Radiation-Reaction problem in General
Relativity'' \cite{10}, which is a good testimony of the
widespread conceptual confusion regarding the GW's. I am grateful
to Prof. G. Morpurgo, who has called my attention on Kennefick's
paper.)

\end{document}